\documentclass{elsart}

\usepackage{graphicx}
\usepackage{amssymb}

\def\NIMA#1#2#3{Nucl. Instr. Meth. Phys. Res. A {\bf #1}\ (#2)\ #3}
\def\NIM#1#2#3{Nucl. Instr. Meth. {\bf #1}\ (#2)\ #3}
\def\IEEE#1#2#3{IEEE Trans. Nucl. Sci. vol. {\bf #1}\ (#2)\ #3}

\begin{document}

\begin{frontmatter}

\title{Space charge in drift chambers operated with the Xe,CO$_2$(15\%) 
mixture}

\author[gsi]{A.~Andronic\thanksref{info}\thanksref{leave}},
\author[hei]{H.~Appelsh{\"a}user}, 
\author[gsi]{C.~Blume}, 
\author[gsi]{P.~Braun-Munzinger}, 
\author[mue]{D.~Bucher}, 
\author[gsi]{O.~Busch}, 
\author[buc,hei]{V.~C{\u a}t{\u a}nescu}, 
\author[buc,gsi]{M.~Ciobanu}, 
\author[gsi]{H.~Daues}, 
\author[hei]{D.~Emschermann}, 
\author[dub]{O.~Fateev},
\author[gsi]{Y.~Foka}, 
\author[gsi]{C.~Garabatos}, 
\author[tok]{T.~Gunji}, 
\author[hei]{N.~Herrmann}, 
\author[tok]{M.~Inuzuka}, 
\author[dub]{E.~Kislov},
\author[kip]{V.~Lindenstruth},
\author[hei]{W.~Ludolphs}, 
\author[hei]{T.~Mahmoud}, 
\author[hei]{V.~Petracek}, 
\author[buc]{M.~Petrovici},
\author[hei]{I.~Rusanov}, 
\author[gsi]{A.~Sandoval},
\author[mue]{R.~Santo},
\author[hei]{R.~Schicker},  
\author[gsi]{R.S.~Simon}, 
\author[dub]{L.~Smykov},
\author[hei]{H.K.~Soltveit}, 
\author[hei]{J.~Stachel}, 
\author[gsi]{H.~Stelzer}, 
\author[gsi]{G.~Tsiledakis}, 
\author[hei]{B.~Vulpescu}, 
\author[mue]{J.P.~Wessels}, 
\author[hei]{B.~Windelband},
\author[hei]{C.~Xu},
\author[mue]{O.~Zaudtke},
\author[dub]{Yu.~Zanevsky},
\author[dub]{V.~Yurevich}

\address[gsi]{Gesellschaft f{\"u}r Schwerionenforschung, Darmstadt, Germany}
\address[hei]{Physikaliches Institut der Universit{\"a}t Heidelberg, Germany}  
\address[mue]{Institut f{\"u}r Kernphysik, Universit{\"a}t M{\"u}nster, Germany}
\address[buc]{NIPNE Bucharest, Romania}  
\address[dub]{JINR Dubna, Russia}  
\address[tok]{University of  Tokyo, Japan}  
\address[kip]{Kirchhoff-Institut f\"ur Physik, Heidelberg, Germany}

{for the ALICE Collaboration}

\thanks[info]{Corresponding author: GSI, Planckstr. 1, 64291 Darmstadt,
Germany; Email:~A.Andronic@gsi.de; Phone: +49 615971 2769; 
Fax: +49 615971 2989.}
\thanks[leave]{On leave from NIPNE Bucharest, Romania. }

\begin{abstract}
Using prototype modules of the ALICE Transition Radiation Detector
we investigate space charge effects and the dependence of the pion
rejection performance on the incident angle of the ionizing particle.
The average pulse height distributions in the drift chambers operated 
with the Xe,CO$_2$(15\%) mixture provide quantitative information on the
gas gain reduction due to space charge accumulating during the drift 
of the primary ionization.
Our results demonstrate that the pion rejection performance of a TRD 
is better for tracks which are not at normal incidence to the anode wires.
We present detailed simulations of detector signals, which reproduce
the measurements and lend strong support to our interpretation of the 
measurements in terms of space charge effects.

\end{abstract}


\end{frontmatter}

\section{Introduction} \label{g:intro}

The ALICE Transition Radiation Detector (TRD) \cite{aa:tdr} is designed
to provide electron identification and particle tracking in the
high-multiplicity environment of Pb+Pb collisions at the LHC.
To achieve the challenging goals of the detector, accurate pulse height 
measurement in the drift chambers operated with a Xe,CO$_2$(15\%) gas 
mixture over the full drift time of the order of 2~$\mu$s is a necessary 
requirement.
For such precision measurements, it is of particular importance first 
to collect \cite{aa:att}, and then to properly amplify all the charge 
created in the detector.
For electrons, the transition radiation, superimposed on the ionization 
charge, provides the required electron/pion identification capability.

For any detector with gas amplification, the positive ions created in 
the avalanche move slowly away from the anode, and this space charge 
leads to a local reduction of the electric field in the proximity of 
the anode.
The effect was recognized in the early days of the development of 
proportional counters and was studied quantitatively later on 
\cite{aa:hend,aa:sip}. In case of multi-wire proportional drift chambers,
the space charge was also studied in detail 
\cite{aa:bres,aa:sauli,aa:boie,aa:emi}, 
as was, more recently, done for electromagnetic calorimeters \cite{aa:na48}.
Several theoretical treatments of the problem have been published 
\cite{aa:hend,aa:sip2,aa:math}.
All of the previous studies concentrated on the resulting gain drop of 
the detector, with its associated loss of efficiency \cite{aa:bres},
as a function of the rate of incoming radiation.
Recently, the effect of space charge on the position resolution 
of drift tubes was also investigated \cite{aa:atlas,aa:atlas2}.

The most obvious impediment caused by space charge concerns the high-rate 
performance of drift chambers (and gaseous detectors in general).
However, space charge can also influence the signal amplitude within a 
single track \cite{aa:bres,aa:emi,aa:det}.
For usual field values applied in multi-wire drift chambers, it takes several 
$\mu$s for the positive ions to move away from the anode surface to a 
distance of several tens of the wire radius \cite{aa:blum}, where their 
effect on the anode field may be considered negligible.
Since in usual drift chambers the arrival time of the primary electrons 
is also of the order of microseconds, the signal will be affected by the 
space charge created by the amplification of primary electrons over the 
full signal collection time (track length).
As noted earlier \cite{aa:bres}, the effect is most pronounced for tracks 
at normal incidence to the anode wires, for which all charge collection 
takes place in a very confined region on the anode wire.
The dimension of an avalanche created by one electron is below 100~$\mu$m, 
essentially independent of its total charge \cite{aa:groh}.
The most significant contribution to the extension of the avalanche for
a cluster of electrons is in fact the spread of the initial ionization due 
to transverse diffusion, which is, for instance, 440~$\mu$m FWHM for 
1~cm drift in our detectors.

We report on measurements of space charge effects within a single track 
in drift chambers operated with the Xe,CO$_2$(15\%) gas mixture.
The measurements were performed during prototype tests for the ALICE TRD. 
The experimental setup and method of data analysis are described in the 
next section. We then present our measurements of the average 
pulse height dependence on drift time as a function of incident angle 
and gas gain. The implications of space charge on electron/pion 
identification are discussed. 
The measurements are compared to simulations, which strongly support our 
interpretation of the results in terms of space charge effects.

\section{Experimental setup} \label{aa:meth} %and method of analysis

\begin{figure}[hbt]
\vspace{.5cm}
%trd_prin3a.fig
\centering\includegraphics[width=.5\textwidth]{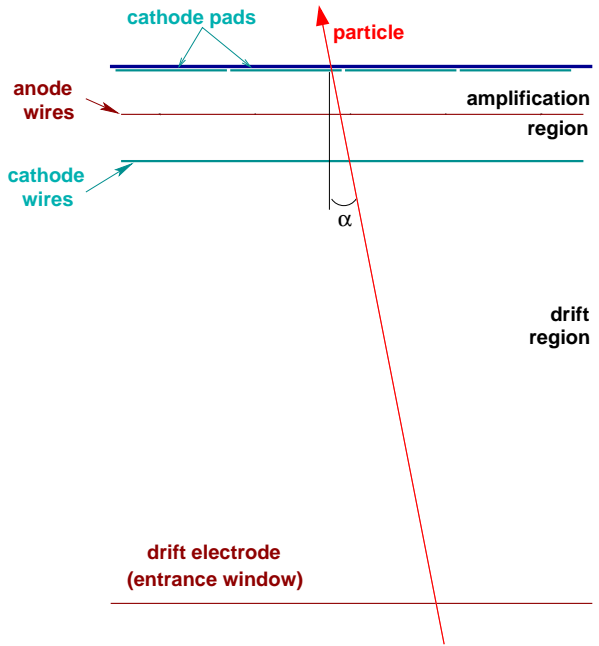}
\caption{Cross-section of a drift chamber along the wire direction.}
\label{g:prin} 
\end{figure}

The results are obtained using prototype drift chambers (DC) with a 
construction similar to that anticipated for the final ALICE TRD 
\cite{aa:tdr}, but with a smaller active area (25$\times$32~cm$^2$).
In Fig.~\ref{g:prin} we present a schematic view of the DC.
As the final detectors for ALICE TRD \cite{aa:tdr}, our prototypes
have a drift region of 30~mm and an amplification region of 7~mm.
Anode wires (W-Au) of 20~$\mu$m diameter are used, with a pitch of 5~mm.
The cathode wires (Cu-Be) have 75~$\mu$m diameter and a pitch of 2.5~mm,
in a staggered geometry.
We read out the signal on a segmented cathode plane with rectangular pads 
of 8~cm length and 0.75~cm width (along the direction of the wires).
The entrance window (25~$\mu$m aluminized Kapton) simultaneously serves  
as gas barrier and as drift electrode.
We operate the drift chambers with the standard gas mixture for the TRD, 
Xe,CO$_2$(15\%), at atmospheric pressure.
The gas is recirculated using a dedicated gas system.

\begin{figure}[hbt]
\vspace{.5cm}
%trd#exe-fe run=-73 |~/mbs/fadc3
\centering\includegraphics[width=.6\textwidth]{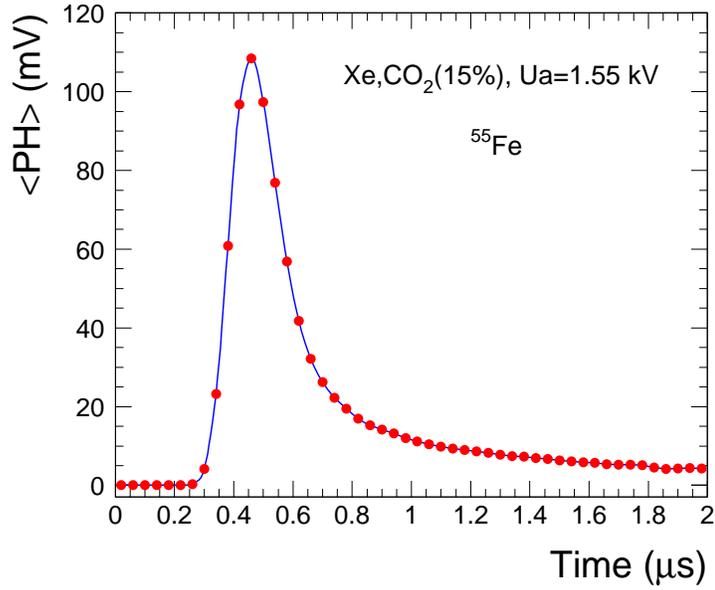}
\caption{Time dependence of the average PASA pulse height for $^{55}$Fe
X-rays.}
\label{g:fe} 
\end{figure} 

We use a prototype of the charge-sensitive preamplifier/shaper (PASA) 
especially designed and built for the TRD in AMS 0.35~$\mu$m CMOS 
technology. 
It has a noise on-detector of about 1000 electrons r.m.s. for our
cathode pad capacitance of about 20~pF (including a 3~pF contribution
of the connection cable).
The FWHM of the slightly asymmetric output pulse is about 100~ns for an 
input step function.
In Fig.~\ref{g:fe} we show an averaged PASA signal for $^{55}$Fe source
X-rays of 5.9~keV.
The signal induced on the pads is determined mainly by the ions moving 
slowly away from the anode, leading to a wider signal compared to the 
intrinsic PASA shaping and to long tails.
The contribution of the longitudinal diffusion to the signal width is 
about 50~ns FWHM.
This convolution of detector signal and PASA response is the so-called 
time response function, TRF.
Here and in the following, the time zero is arbitrarily shifted by about 
0.3~$\mu$s to facilitate a simultaneous measurement of the baseline and 
of noise.
The nominal gain of the PASA is 12~mV/fC, but during the present measurements 
we lowered the gain to 6~mV/fC for a better match to the range of our 
Flash ADC (FADC) system with 0.6~V voltage swing.
The FADC used for the tests is different from the one designed for 
the final detector \cite{aa:tdr}.
It has an 8-bit non-linear conversion and adjustable baseline,
running at 20~MHz sampling frequency.
The data acquisition (DAQ) is based on a VME event builder and was developed 
at GSI \cite{aa:mbs}. % RIO2 \cite{aa:rio}.
As the beam diameter is of the order of 3~cm FWHM, we usually limit the 
readout of the DC to 8 adjacent pads. This also minimizes data transfer 
on the VSB bus connecting the FADC and the event builder.

Four identical layers of radiator and drift chamber were used for these 
measurements.
The variation of the gas gain for each individual chamber is within 10\%.
The results presented below are averaged over these four drift chambers 
in order to improve statistics.
The radiator used for these measurements is of the same design as 
the one envisaged for the final ALICE TRD \cite{aa:tdr}. 
It is a sandwich of Rohacell foam and polypropylene fibres. 
A reinforcement of carbon fibres of about 100~$\mu$m thickness is applied 
to the outer surfaces of this sandwich to ensure for the real-size detectors
the flatness of the drift electrode, which will be directly glued on the 
radiator, for overpressures up to 1~mbar.

To study the effect of space charge on the time evolution of the average
signal, we vary the angle of incidence of the beam with respect to the 
normal incidence to the anode wires.
A particle trajectory through the detector is sketched in Fig.~\ref{g:prin}.
The measurements were carried out at the T10 secondary beamline of the 
CERN PS \cite{aa:cernpi} at the beam momentum of 3~GeV/c.
The resolution of the beam momentum is $\Delta p/p\simeq 1\%$.
The beam intensity was up to 3000 particles per spill of about half a second.
The beam contains a mixture of electrons and negative pions, with an 
electron content of about 2\%.
Similar sample sizes of pion and electron events were acquired under exactly 
the same detector conditions, via dedicated triggers.
For the present analysis we have selected clean samples of pions and electrons
using coincident thresholds on two Cherenkov detectors and on a lead-glass 
calorimeter \cite{aa:andr}.

\section{Experimental results} \label{g:xe}

We performed the measurements for four values of gas gain, 2400, 3900, 6200 
and 9600, corresponding to anode voltages of 1.5, 1.55, 1.6 and 1.65 kV.
The drift field for the nominal drift voltage of -2.1~kV varies from 725 to 
733~V/cm for our range of the anode voltages.
The incident angle with respect to the normal to the pad plane was varied
from 0$^\circ$ to 15$^\circ$ in steps of 5$^\circ$ by tilting the detectors 
with respect to the direction of the beam about an axis perpendicular to 
the wires (see Fig.~\ref{g:prin}) and perpendicular to the drift direction.

\begin{figure}[htb]
\vspace{-.5cm}
%trd#pl-ang
\centering\includegraphics[width=.65\textwidth]{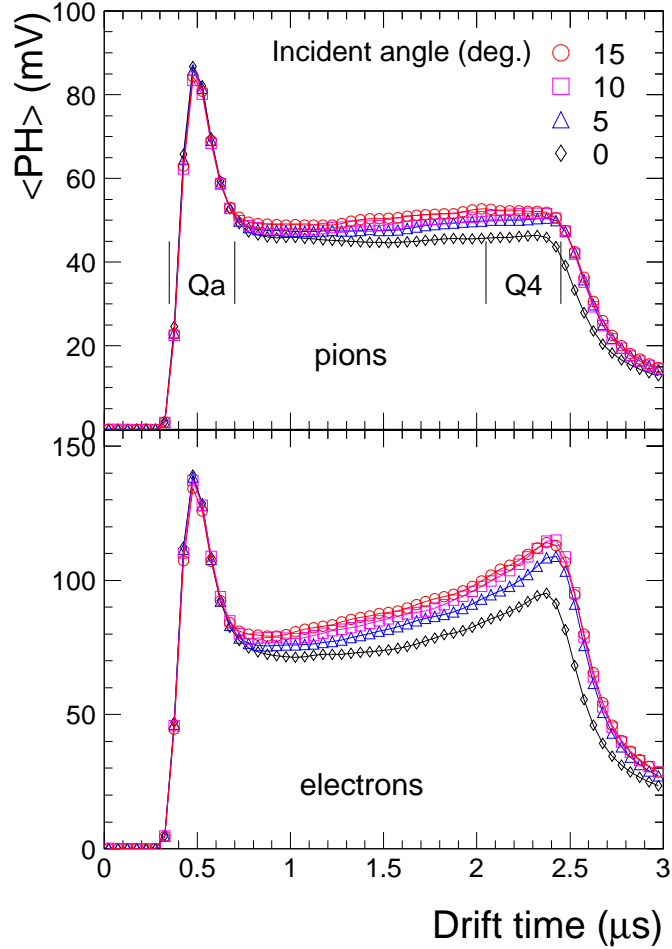}
\caption{Average pulse height as a function of drift time, for different 
incident angles. Upper panel: pions, lower panel: electrons. The gas gain 
used for these measurements was 3900.}
\label{g:xe1}
\end{figure}

In Fig.~\ref{g:xe1} we present the dependence of the average pulse height, 
$\langle PH \rangle$, on the drift time, for different incident 
angles $\alpha$, for pions and electrons.
These average signals were recorded for the anode voltage of 1.55~kV.  
They are the sum of all eight pads and consequently are not influenced by
the charge sharing between adjacent pads, which depends on the angle of 
incidence.
The overlap of the long ion tails (TRF) results, in case of pions, in a 
slightly rising average pulse height as a function of the drift time, 
as seen in Fig.~\ref{g:xe1} for large angles.
The peak at short drift times originates from the primary ionization 
(d$E$/d$x$) generated in the amplification region, where the signal from 
both sides of the anode wires overlaps in the same time interval. 
Due to the stronger field, the drift velocity in the amplification region 
is considerably larger than in the drift region.
Consequently, the timing characteristics of the signal from the amplification 
region is determined mainly by the shaping time of the PASA.
As the angle decreases towards normal incidence, the signal is progressively 
attenuated as a function of drift time, a clear indication of the 
effect of space charge.
As this is a local effect, when spreading the primary electrons along 
the anode wire, the effect becomes less important, at least for our moderate 
values of the gain.
Note that the amplitude of the signal in the amplification region is 
independent of the angle, since there are no previous avalanches that 
can screen it.
Only a trivial normalization of the data for different angles is 
done to take into account the variation of the effective track length 
with the angle.
We note that our measurements established for the first time 
\cite{aa:andr,aa:att} the expected time evolution of the signal in this 
type of drift chambers.
It is possible that earlier measurements \cite{aa:wat,aa:app,aa:det},
showing a decreasing value of the average signal in the drift region,
suffered from space charge effects within a single track, due to the normal
incidence used.
For electrons, the contribution of transition radiation (TR), which is
absorbed preferentially at the entrance of the drift chamber and is 
registered superposed on d$E$/d$x$, results in the strong rise of the average 
signal towards the end of the drift time.
The d$E$/d$x$ of electrons is in the regime of the Fermi plateau and 
consequently is on average about 40\% larger than for pions at 3~GeV/c 
\cite{aa:dedx}.

\begin{figure}[htb]
\vspace{-.5cm}
%trd#pl-ang pl=qra2
\centering\includegraphics[width=.63\textwidth]{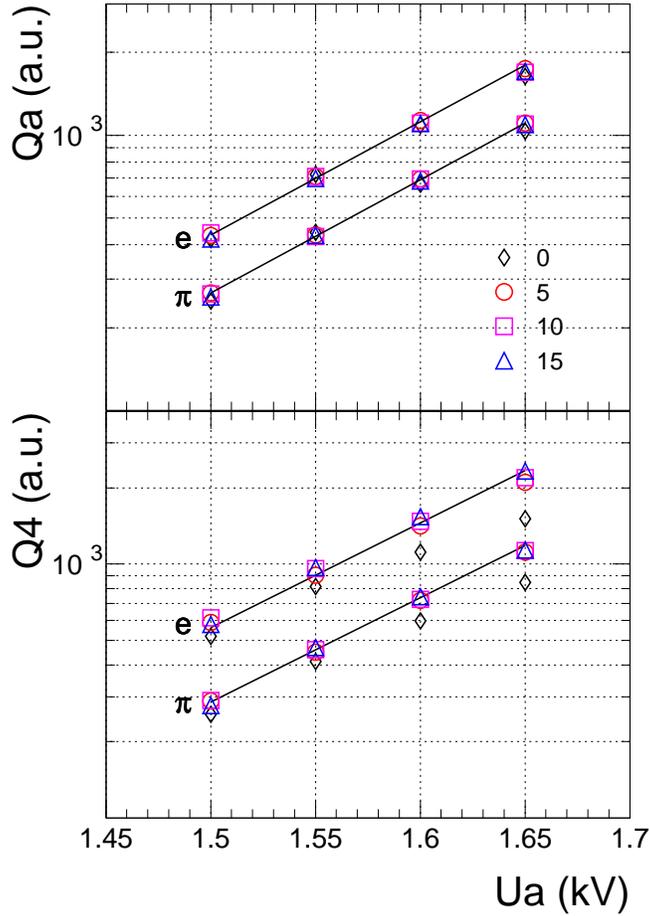}
\caption{The dependence of collected charges Qa and Q4 (see text) on the 
anode voltage, for pions and electrons.}
\label{g:xe1b}
\end{figure}

Marked in Fig.~\ref{g:xe1} by vertical lines are the limits used to
calculate the average integrated charges in the amplification 
region, Qa, and in the last quarter of the drift region, Q4.
These charges are plotted in Fig.~\ref{g:xe1b} as a function of the
anode voltage for all the incident angles, for pions and electrons.
These dependences reflect the exponential gas gain increase as a function
of the anode voltage, represented by the lines.
In the presence of space charge the measured charges would flatten for 
higher anode voltages \cite{aa:blum}, in particular at normal incidence, 
when the charge is collected in a narrow spot on the anode wire.
An exponential behavior is seen for Qa for all angles, demonstrating
that no space charge effects due to rate occur for our voltage values.
Taking into account our beam conditions (3000 particles in half a second,
spread uniformly in a disk of 3 cm diameter) results in a rate of about 
50~Hz per mm of anode wire. This is a very low local rate, compared to
the value of 10$^6$~Hz/mm, estimated from the X-rays measurements of 
ref.~\cite{aa:boie} to be the onset of rate-induced space charge effects 
for a gas gain of about 4000.
The fact that only Q4 shows a flatter dependence on anode voltage for 
the normal incidence can only be caused by the space charge produced by
the avalanches of the earlier ionization electrons of the same track.
Larger d$E$/d$x$ and the contribution of TR makes the effect larger for 
electrons.

\begin{figure}[htb]
\vspace{-.5cm}
%trd#pl-ang pl=qra2
\centering\includegraphics[width=.65\textwidth]{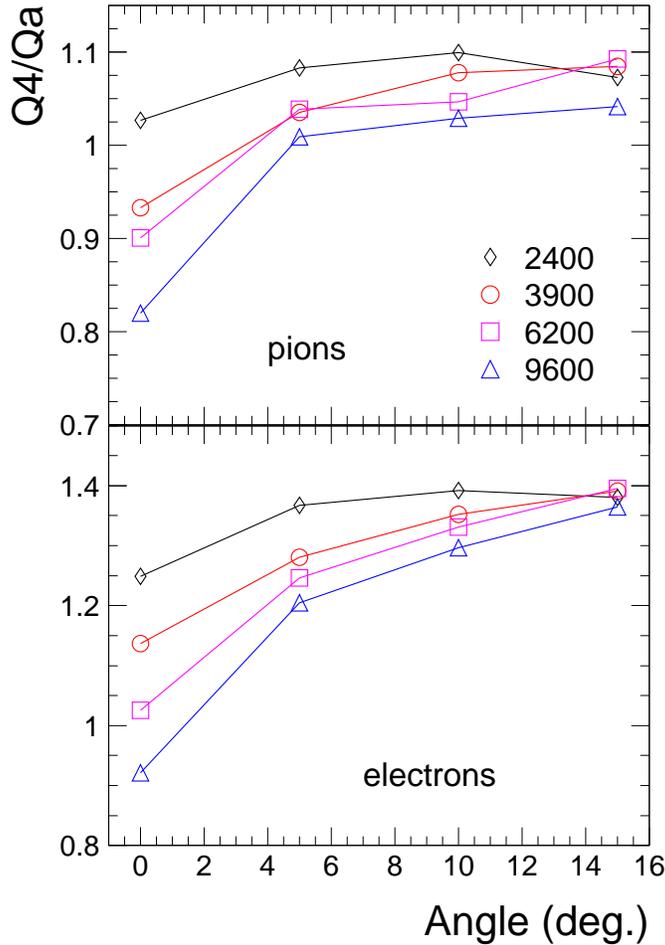}
\caption{Ratios of charges recorded in the drift and amplification
region as a function of incident angle for four values of the gas gain,
for pions and electrons.}
\label{g:xe2}
\end{figure}

The ratio Q4/Qa is plotted in Fig.~\ref{g:xe2} as a function of the 
incident angle for four values of the gas gain for pions and electrons. 
For pions, due to track segment considerations, this ratio should be close 
to unity, but its absolute value is influenced by the integration windows
(through the finite time bin size) and by the TRF.
For electrons, due to the contribution of TR, the ratio has a larger 
value. In the absence of screening due to space charge, for both pions 
and electrons this ratio would be independent of the incident angle, 
but we observe a marked variation as a function of angle, in particular 
a sharp drop for small angles. 
A saturation is reached at large angles due to the locality of the screening.
This behavior was observed before with an Ar-based mixture \cite{aa:bres}, 
albeit with a different magnitude, due probably to the much larger gas 
gain used in that study.
More recently, similar results were obtained for a He-based mixture 
\cite{aa:emi}.
As expected from space charge considerations, we observe a stronger 
variation of the ratio for larger gains, in qualitative agreement with 
other observations \cite{aa:emi}.

\begin{figure}[htb]
\vspace{-.5cm}
%trd#pl-ang pl=rphepi |mbs/fadc4
\centering\includegraphics[width=.8\textwidth]{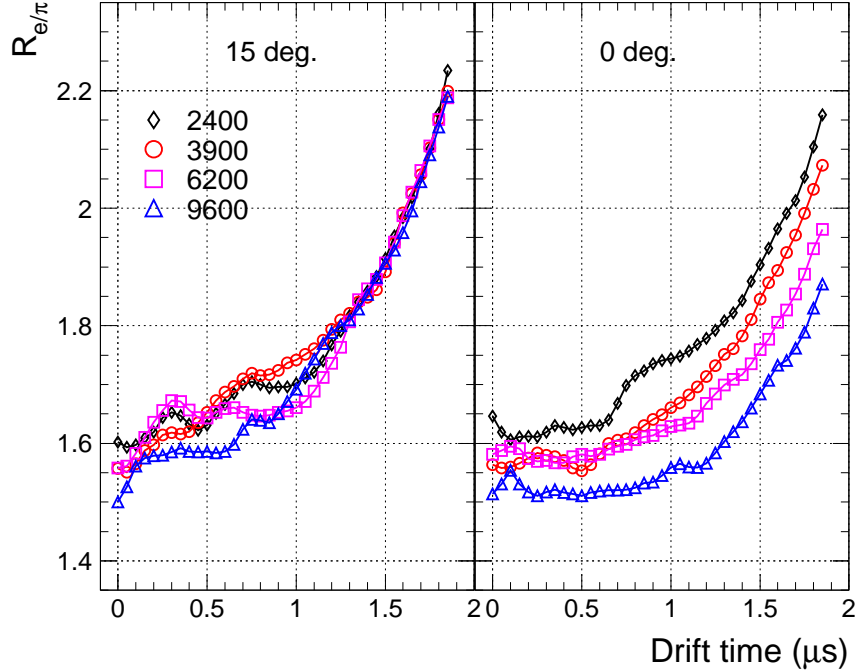}
\caption{Ratio of signal of electrons and pions as a function of drift time 
for four values of the gas gain. Left panel is for 15$^\circ$ incidence,
right panel is for 0$^\circ$ incidence.}
\label{g:xe4}
\end{figure}

In Fig.~\ref{g:xe4} we present the ratio of the average signal of electrons 
to pions, $R_{e/\pi}$, as a function of drift time, for two extreme cases of 
incidence, 15$^\circ$ and 0$^\circ$ and for four values of the gas gain.
The time dependence of $R_{e/\pi}$ is due to the contribution of TR.
This ratio is independent of the gas gain for the angle of 15$^\circ$, 
when space charge plays no role. Conversely, at normal incidence, when space 
charge is most important, a progressive reduction of $R_{e/\pi}$ is seen 
as a function of gas gain.
The ratio $R_{e/\pi}$ is a direct measure of the electron/pion separation power
of a TRD.

\begin{figure}[hbt]
\vspace{-.5cm}
%trd#eff-ang eff=3
\centering\includegraphics[width=.75\textwidth]{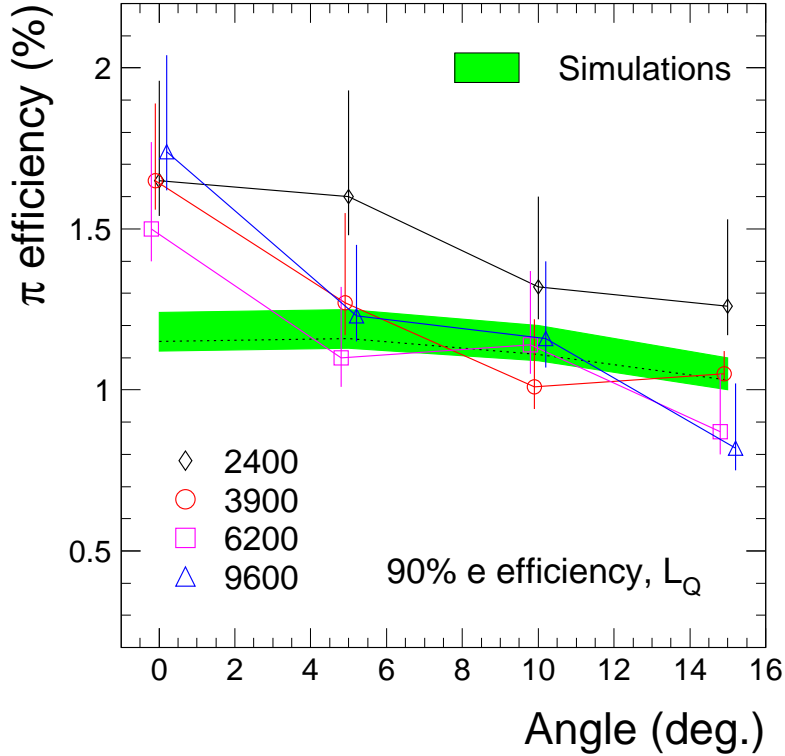}
\caption{Pion rejection performance as a function of incident angle for
a 6 layer TRD.
The symbols represent data. The shaded area shows results of simulations, 
without inclusion of space charge effects.} 
\label{g:xe5}
\end{figure}

The observed stronger attenuation of the signal due to space charge for 
electrons compared to pions does affect the electron/pion identification 
performance of a TRD. 
For our case, we calculate the pion efficiency for 90\% electron efficiency
using a likelihood method \cite{aa:bun} on the total energy deposited in 
a single chamber. 
Again, to improve statistics, each of the four layers has been treated 
as a separate detector and its total charge filled in a common histogram, 
for pions and electrons separately.
We use these two charge distributions in a simulation program to calculate
the likelihood (to be an electron) for a six layer detector, corresponding 
to the configuration of ALICE TRD.
A 90\% efficiency cut has been selected in the likelihood distribution
of electrons and the pion contamination has been calculated.
The results are presented in Fig.~\ref{g:xe5} as a function of incident 
angle for our four values of the gas gain.
A pion rejection factor (inverse of the pion efficiency) of about 100 is 
achieved for finite angles of incidence, fulfilling the ALICE design goal 
\cite{aa:tdr}.
One can see the expected degradation of the pion rejection power as the 
incident angle approaches normal incidence.
Besides the space charge, two other contributions have to be recognized:
i) the improvement of pion rejection for larger gains due to larger values 
of the signal-to-noise ratio (S/N); 
ii) the improvement of pion rejection as a function of angle arising from 
a thicker effective radiator and detector.
To quantify this last effect, we calculate the angle dependence of the 
pion efficiency using simulated events. A parametrization for a regular 
radiator was tuned to reproduce the measured pion efficiencies at 15$^\circ$. 
Space charge effects are not included in the simulations, which are performed 
at a constant S/N. 
The results are represented by the dotted line and the shaded area in 
Fig.~\ref{g:xe5}.
Although the statistical errors of the measurements are rather large, 
an obvious degradation of the pion rejection is observed for normal 
incidence.
For the upper values of our gas gain we measure at $0^\circ$ a pion 
rejection worse by a factor of 1.5 beyond the expected contribution due 
to the effective thicknesses.
A similar degradation for normal incidence was observed, albeit with a 
much greater magnitude, in electron/pion identification using d$E$/d$x$ 
measurements with prototypes for the PHENIX experiment
and was also attributed to space charge \cite{aa:libby}.

Since our FADC measurements make available the time dependence of
the avalanche charge for each individual track, it is conceivable that 
the space charge effects can be corrected using this information.
We have implemented such a procedure, in which the measured amplitude 
of a given time bin is corrected by a factor which depends on the total 
charge registered prior to this time bin. %, $Q_{prev}$.
The correction is done event-by-event, in an identical ("blind") way 
for both pions and electrons and is tuned to restore the average ratio
R$_{\langle PH \rangle}$ (see Fig.~\ref{g:xe3}) to a flat dependence
on the drift time.
This correction is successful in restoring the ratio $R_{e/\pi}$ 
(see Fig.~\ref{g:xe4}) for 0$^\circ$ incidence to the value measured 
at 15$^\circ$. However, the pion rejection factor is improved only 
marginally because the fluctuations of the charge distributions
are amplified by the correction. For instance, the r.m.s. of the pion charge
distribution is 77\% before and 87\% after correction (compared to 71\%
at 15$^\circ$ incidence).

The degradation of the electron/pion identification performance for tracks 
approaching normal incidence to the anode wires is an important argument 
for operating detectors at the lowest possible gas gain.
Concerning the ALICE TRD, the optimal gas gain value is a compromise
between the requirements on pion rejection and on position resolution, 
which is strongly improving as a function of S/N.
We note that, due to the geometry of the ALICE TRD, normal incidence 
occurs rather seldom.
The features presented above in case of electron/pion identification 
with TRD apply also to the identification of other particle species
using energy loss in ionization detectors. 
Lighter gas mixtures show the effect of space charge as well 
\cite{aa:emi,aa:hau}. 
%although with a lower magnitude as compared to Xe-based mixtures at the same 
%gas gain, due to the larger mobility of lighter ions \cite{aa:blum}.
%However, due to lower primary ionization, for a given S/N value, lighter 
%gas mixtures have to be operated at larger gains, which would lead to 
%roughly similar space charge effects. 

\section{Comparison to simulations} \label{aa:sim}

For a quantitative understanding of the observations presented before 
we have performed a Monte Carlo simulation of the detector signal. 
The underlying physical picture has been described in \cite{aa:blum}
and will be briefly summarized in the following.

The electric field around the anode wire is assumed to correspond to an
ideal cylindrical geometry and is given by the charge density $\lambda$ 
on the wire:
\begin{equation}
	E_0(r) = \frac{\lambda}{2\pi\epsilon_0 r}
	\label{efield}
\end{equation}
with $\lambda=\sigma V$, where $\sigma$ is the wire capacitance per 
unit length and $V$ the anode voltage.
The ions produced in an avalanche form a cylinder of positive space-charge 
with radius $R$ and charge density $\lambda^{\star}$ surrounding the anode 
wire. The presence of the space-charge cylinder leads to a modification
of the charge density on the wire.
Inside the space-charge cylinder the electric field is
\begin{equation}
 	E_{\rm in}(r) = \frac{\lambda'}{2\pi\epsilon_0 r}~~~~~~~ (r<R)
\end{equation}
where $\lambda'$ is the modified charge density on the wire. Outside
the space-charge cylinder the field is
\begin{equation}
 E_{\rm out}(r) = \frac{\lambda' + \lambda^{\star}}{2\pi\epsilon_0 r} 
 \qquad (r>R).
\end{equation}
The value of $\lambda^{\star}$ is determined by the potential difference
$V$ between anode wire and cathode:
\begin{equation}
 V= \frac{\lambda}{2\pi\epsilon_0}\int_{a}^{b}\frac{{\rm d}r}{r} 
  = \frac{\lambda'}{2\pi\epsilon_0}\int_{a}^{R}\frac{{\rm d}r}{r}
  + \frac{\lambda'+\lambda^{\star}}{2\pi\epsilon_0}\int_{R}^{b}
    \frac{{\rm d}r}{r}
 \label{potential}
\end{equation}

\begin{equation}
 \Rightarrow \lambda' = \lambda - \lambda^{\star} \frac{\ln b/R}{\ln b/a},
\end{equation}
with anode wire radius $a$ and anode-cathode distance $b$.
The charge density on the wire is reduced by
\begin{equation}
  \frac{{\rm d}\lambda}{\lambda} = \frac{\lambda - \lambda'}{\lambda} 
	= \eta(T)\frac{\lambda^{\star}}{\lambda},
	\label{dlambda}
\end{equation}
where 
\begin{equation}
\eta(T) = \frac{\ln b/R(T)}{\ln b/a}.
\end{equation}
The shielding factor $\eta(T)$ decreases with increasing $R$ and thus
depends on the drift time $T$ of the ion cloud because the ions are slowly 
drifting towards the surrounding cathodes.
The drift time $T$ is determined by the ion mobility $\mu$
\begin{equation}
  T= \int_{a}^{R} \frac{{\rm d} r}{\mu E(r)} = \frac{R^2 - a^2}{2a\mu E(a)},
\end{equation}
assuming that the ion drift starts at the wire surface at $t=0$. 

The relative gain variation ${\rm d} G/G$ depends on the variation of the
charge density ${\rm d}\lambda/\lambda$ via~\cite{aa:blum}:
\begin{equation}
\frac{{\rm d}G}{G} = 
\left( \ln G + \frac{\lambda\ln 2}{\Delta V 2\pi\epsilon_0}\right)
\frac{{\rm d}\lambda}{\lambda},
\label{dg}
\end{equation}
with the Diethorn parameter $\Delta V \approx 30$~V in xenon mixtures.

In the case of the TRD, the multiplication of a given drift electron is reduced
by the shielding effect of the ion clouds produced by preceding electrons in 
the same event. As input to the simulation we use the spatial distribution of 
ionization electrons along the particle trajectory in the detector gas. 
The arrival time at the anode wire has been calculated using GARFIELD. 
The arrival point along the wire is determined by the incident angle of 
the simulated track and smeared by transverse diffusion.

The computation of the actual multiplication factor $G_i$ of an electron $i$ 
requires the consideration of the ion clouds with charge density 
$\lambda^{\star}_j$ built up by previous electrons $j$ ($j<i$)
and their respective shielding factors $\eta_j(T_j)$. The ion drift time 
$T_j$ is equal to the difference of the arrival times $t_i-t_j$ of the 
electrons $i$ and $j$.
The ion charge density $\lambda^{\star}_j$ produced by electron $j$ with 
elementary charge $e$ is given by
\begin{equation}
 \lambda_j^{\star} = e \cdot G_j/L,
\end{equation}
where $L$ is the lateral extent of the avalanche along the wire and $G_j$ 
is the actual multiplication factor of electron $j$ which itself had been 
reduced by previous electrons.

Using Eq. (\ref{potential}) the modified charge density $\lambda'_i$ at the 
time of the arrival of electron $i$ can be 
calculated:
\begin{equation}
  V= \frac{\lambda}{2\pi\epsilon_0}\int_{a}^{b}\frac{{\rm d}r}{r} \\ \nonumber
   = \frac{\lambda'_i}{2\pi\epsilon_0}\int_{a}^{R_1}\frac{{\rm d}r}{r}
     + \frac{\lambda'_i+\lambda^{\star}_1}{2\pi\epsilon_0}\int_{R_1}^{R_2}\frac{{\rm d}r}{r}
     + \dots 
     + \frac{\lambda'_i+\lambda^{\star}_1 + \dots + \lambda^{\star}_{i-1}}{2\pi\epsilon_0}
	\int_{R_{i-1}}^{b}\frac{{\rm d}r}{r}. 
   \label{potall}
\end{equation}
For the reduction of the charge density at the arrival time of electron $i$ 
we obtain:
\begin{equation}
  \frac{{\rm d}\lambda_i}{\lambda} = \frac{\lambda-\lambda'_i}{\lambda}
	= \frac{1}{\lambda} \sum_{j=1}^{i-1} \lambda^{\star}_j \eta_j(T_j)
   \label{dlambdaall}
\end{equation}
which is identical to Eq.~(\ref{dlambda}) for the case $i=2$.
Inserting this result into Eq.~(\ref{dg}) yields the multiplication
factor $G_i = G(1-{\rm d}G_i/G)$ for electron $i$.

It is important to note that the contribution of an electron $j$ in 
Eq.~(\ref{potall}) and (\ref{dlambdaall}) is only considered if the lateral 
distance of the arrival points at the wire of electrons $i$ and $j$ is smaller
than $L/2$. In this way, the dependence of the space charge effect on the
incident angle is introduced. It is also required that the ions have 
drifted at least 50~$\mu$m away from the wire to ensure that the ions
are outside the amplification region. 

The number of primary electrons used in the simulations are the measured
values.
For our momentum of 3~GeV/c, we measure \cite{aa:dedx} an average energy 
deposit for pions of 5.5 keV/cm, which amounts for our gas mixture to 
243 primary electrons per cm.
170 electrons from the amplification region arrive at the anode in a fraction 
of a $\mu$s, while, from the drift region, 729 electrons arrive at a constant 
rate over the drift time of 1.8~$\mu$s. 
For instance, after gas amplification this corresponds, for a gas gain 
of 3900, to a total charge of 106~fC and 455~fC, respectively.
For electrons we use the time-dependent ratio measured at 15$^\circ$ incidence
(left panel of Fig.~\ref{g:xe4}), which is independent of the gas gain.
All the other input values used in the calculations are summarized in 
Table~\ref{sim_tab}.

 \begin{table}[htb]
% \centering
\caption{List of the input values used in the simulations.}
\vspace{0.25cm}
\tabcolsep=20pt
\begin{tabular}{lr} 
Parameter & Value \\ \hline
Anode voltage $V$ & 1550 V, 1600 V, 1650 V \\ 
Gas gain $G$ & 3900, 6200, 9600 \\
Anode wire radius $a$ & 10~$\mu$m \\ 
Anode-cathode distance $b$ & 3.5~mm \\ 
Ion mobility $\mu$ & $0.57 \cdot 10^{-6}$~cm$^2$/V/$\mu$s \\ 
$\Delta V$ & 30 V \\
Wire capacity $\sigma$& $9.5\cdot 10^{-14}$ F/cm \\ 
Avalanche spread $L$ & 50~$\mu$m \\ 
Transverse diffusion coefficient $D_t$ & 330~$\mu$m/$\sqrt{\rm cm}$\\ 
\end{tabular}
\label{sim_tab}
\end{table}

\begin{figure}[htb]
\vspace{-.5cm}
%sim#pl-ang
\centering\includegraphics[width=.75\textwidth]{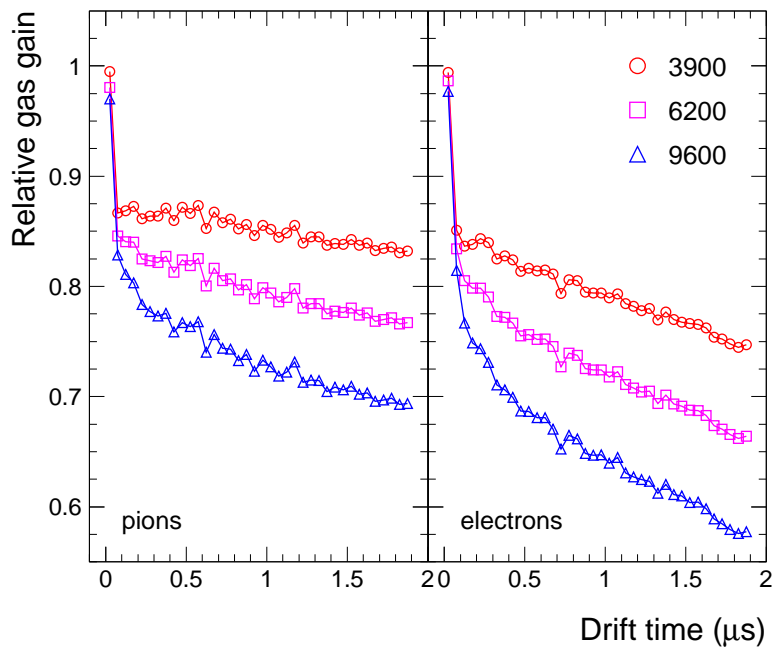}
\caption{Simulated relative gain as a function of drift time at normal 
incidence for three values of the gas gain, for pions and electrons.}
\label{g:gain}
\end{figure}

The calculated average relative gain values as a function of drift time 
are shown in Fig.~\ref{g:gain} for pions and electrons at normal incidence,
for three values of the gas gain.
The sharp drop of the gas gain in the first 0.1~$\mu$s is due to the effect 
of the large charge densities from the energy deposit in the amplification 
region.
After this, the gain reduction due to space charge approximately levels off 
in case of pions as a result of an equilibrium between the incoming charge 
from the drift region at a roughly constant rate and the movement of the 
ions from previous avalanches away from the anode.
The gain reduction for electrons is stronger than for pions and with a more
pronounced time dependence. As already explained, this is due to the larger 
average signals for electrons, in particular with the contribution
of TR at large drift times.
For the largest value of the gas gain a reduction of the signal 
at the end of the drift time by about 30\% is observed for pions
and by about 40\% for electrons.

\begin{figure}[htb]
\vspace{-.5cm}
%sim#pl-ang
\centering\includegraphics[width=.65\textwidth]{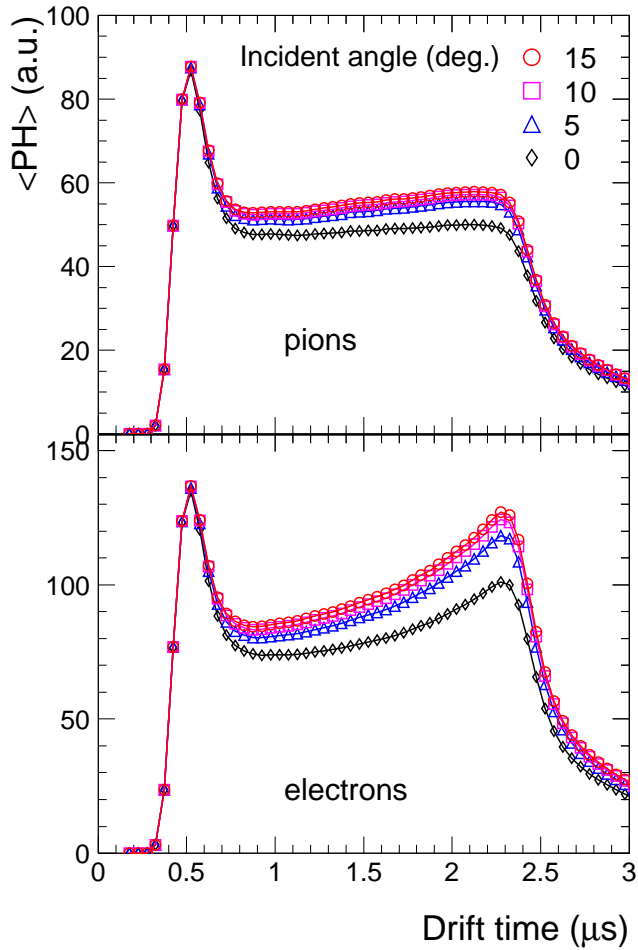}
\caption{Simulated average pulse height as a function of drift time, 
for different incident angles for the gas gain of 3900, for pions and
electrons.}
\label{g:sig}
\end{figure}

For an exact description of the measured signals, the arrival time 
distribution of the primary electrons is folded by the single-electron TRF, 
which has been determined experimentally from the signal shape of 
$^{55}$Fe events (see Fig.~\ref{g:fe}).
The resulting simulated signals are shown in Fig.~\ref{g:sig} for tracks
with different incident angles. 
A very good overall agreement with the measured signals is seen.
As in case of the measurements, a clear reduction of the pulse height 
for tracks with small incident angle can be observed as a function of 
drift time. 

\begin{figure}[htb]
\vspace{-.5cm}
%sim#pl-ang pl=rphxe |mbs/fadc4
\centering\includegraphics[width=.8\textwidth]{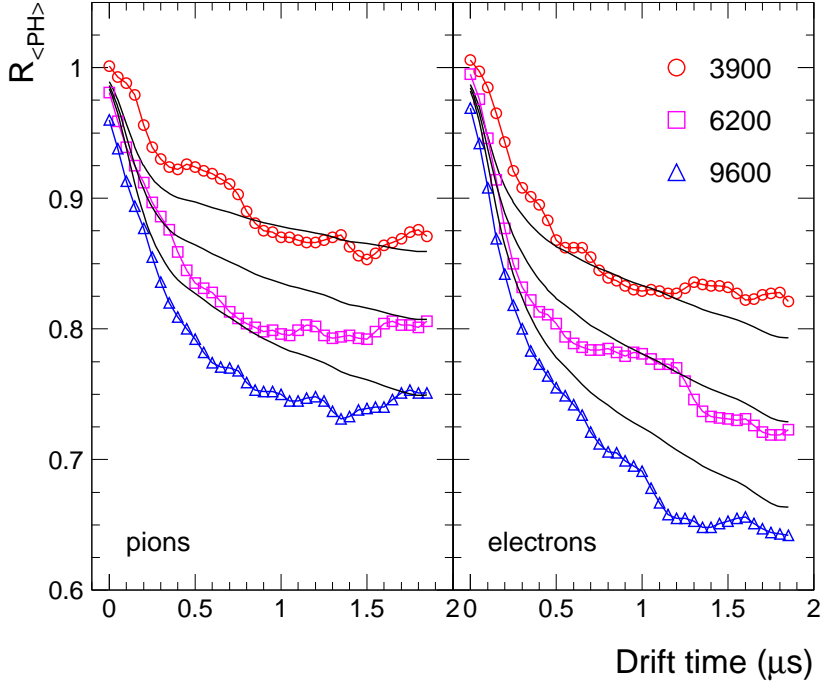}
\caption{Ratios of signals at 0$^\circ$ and 15$^\circ$ incidence as a 
function of drift time for three values of the gas gain, for pions 
and electrons. The measurements (symbols) are compared to calculations
(lines).
}
\label{g:xe3}
\end{figure}

To quantify the signal reduction, we construct the ratio of the average 
signal at normal incidence to the average signal at the largest incidence, 
R$_{\langle PH \rangle}$. 
From the measured behaviour of charge ratios as a function of angle, seen 
in Fig.~\ref{g:xe2}, one can conclude that the screening is negligible for 
our largest angle of incidence. 
Indeed, our simulations show that, for 15$^\circ$ incidence, the gain at 
the end of the drift is reduced by only 1.5\%.
As a consequence, R$_{\langle PH \rangle}$ is a quantitative measure 
of the screening at normal incidence.
In Fig.~\ref{g:xe3} we present the dependence of this ratio on drift time 
for pions and electrons for three values of the gas gain.
The time reference has been chosen to be the time of the maximum 
$\langle PH \rangle$ (corresponding to $t\simeq0.5~\mu$s in Figs.~\ref{g:xe1}
and \ref{g:sig}).
The measurements are compared to calculations. 
The fine structure of the measured data is an artifact of signal fluctuations 
due to limited track statistics.
Despite the simplifying assumptions involved, the calculations are in a 
reasonable agreement with the measurements, in the magnitude as well as 
in the shape.
%hence demonstrating that space charge is the underlying mechanism for the 
%observed signal reduction within a single track.
%Note that already at this "time zero" there is a small screening for 
%the larger gains seen in the data, but not in the calculations.

\section{Summary} \label{aa:sum}

We have reported measurements on space charge effects within a single track
and the dependence of the pion rejection performance on the incident angle, 
carried out using prototype detectors for the ALICE TRD.
Our measurements of average pulse height distributions in drift chambers
operated with Xe,CO$_2$(15\%) provide quantitative results on the signal
reduction within a given track due to space charge accumulation during 
the drift of the primary ionization. 
We have shown that the pion rejection performance of a TRD is impaired 
for tracks at normal incidence to the anode wire plane.
Since in general normal incidence cannot be avoided in drift chambers,
the only possibility to minimize the space charge effects is to chose 
the lowest possible gas gain allowed by a reasonable compromise on the
desired position resolution of the detector.
Our detailed simulations of the detector signals are in a remarkable 
agreement with the measurements, hence demonstrating that space charge 
is the explanation for the observed signal reduction within a signal track 
at normal incidence to the anode wires.

\section*{Acknowledgments}
We acknowledge A.~Radu and J.~Hehner for the skills and dedication in building
our detectors and N.~Kurz for help on data acquisition.
We would also like to acknowledge P. Szymanski for help in organizing the
experiment and A.~Przybyla and M.~Wensveen for technical assistance during 
the measurements.

%\clearpage

\end{document}